\documentclass[12pt]{iopart}

\usepackage{comment}
\usepackage{color, xcolor}
\usepackage[hypertexnames=false, unicode, colorlinks, allcolors=blue!70!black, linktocpage, pdfusetitle]{hyperref}
\usepackage[all]{hypcap}
\usepackage{amssymb,amsbsy}
\usepackage{indentfirst}
\usepackage[pdftex]{graphicx}
\usepackage[numbers,square,comma,sort&compress]{natbib}
\usepackage{url}

\usepackage[T1]{fontenc}
\usepackage[utf8]{inputenc}
\usepackage{lmodern}

\usepackage{etoolbox}

\makeatletter
\newcommand{\mainmatter}{%
  \setcounter{footnote}{0}%
  \patchcmd{\@makefntext}{\fnsymbol}{\arabic}{}{}%
  \patchcmd{\@thefnmark}{\fnsymbol}{\arabic}{}{}%
  \def\@makefnmark{\textsuperscript{\arabic{footnote}}}%
}
\makeatother
\begin{document}

\title{Towards a Non-singular Paradigm of Black Hole Physics}

\author{
Raúl~Carballo-Rubio$^{1,2}$, Francesco~Di~Filippo$^{3}$, Stefano~Liberati$^{4,5}$, Matt~Visser$^{6}$}

\author{Julio Arrechea$^{4,5}$, Carlos~Barceló$^{1}$,
Alfio~Bonanno$^{7,8}$,
Johanna~Borissova$^{9,10}$,
Valentin~Boyanov$^{11}$,
Vitor~Cardoso$^{2,11}$,
Francesco~Del~Porro$^{12}$,
Astrid~Eichhorn$^{13}$,
Daniel~Jampolski$^{3}$,
Prado~Mart\'in-Moruno$^{14}$,
Jacopo~Mazza$^{15}$,
Tyler~McMaken$^{16}$,
Antonio~Panassiti$^{7,8,17}$,
Paolo~Pani$^{18}$,
Alessia~Platania$^{12}$,
Luciano~Rezzolla$^{3,19,20}$,
Vania~Vellucci$^{21}$
}

\begin{abstract} 
The study of regular black holes and black hole mimickers as alternatives to standard black holes has recently gained significant attention, driven both by the need to extend general relativity to describe black hole interiors, and by recent advances in observational technologies. Despite considerable progress in this field, significant challenges remain in identifying and characterizing physically well-motivated classes of regular black holes and black hole mimickers. This paper provides an overview of these challenges, and outlines some of the promising research directions --- as discussed during a week-long focus program held at the Institute for Fundamental Physics of the Universe (IFPU) in Trieste from November 11th to 15th, 2024.
\end{abstract}

\markboth{Towards a Non-singular Paradigm of Black Hole Physics}{}

\maketitle
\tableofcontents

\newpage

\markboth{Towards a Non-singular Paradigm of Black Hole Physics}{}

\section{Introduction}

Karl Schwarzschild found the first non-trivial exact solution to the Einstein field equations~\cite{Schwarzschild:1916dge} just a few months after their publication in 1915~\cite{Einstein1915:paw}. However, it took several decades for the scientific community to fully understand the physical implications of this solution~\cite{Corda:2010ft}. Its interpretation required the development of groundbreaking concepts, such as event horizons and spacetime singularities, which were unprecedented in physics~\cite{Hawking:1973uf}. Thanks to the contributions of multiple generations of researchers, these ideas are now fundamental tools in the standard repertoire of relativists~\cite{Poisson:2009pwt}.

The Schwarzschild solution describes the gravitational field outside a spherical non-rotating mass, which can either have a finite size or be a point particle. In the latter case, the Schwarzschild solution displays a spacetime singularity surrounded by an event horizon. Both elements are the defining characteristics of a \emph{standard black hole},\footnote[1]{We use the adjective \emph{standard} to distinguish the solutions of the Einstein field equations (or a suitable generalization thereof), which are purely mathematical concepts, from the physical concept of a black hole as an astronomical object.} of which the Schwarzschild solution is a representative.

Perhaps surprisingly, both event horizons and spacetime singularities are generally considered to be mathematical idealizations that, while useful approximations to reality, fall short of fully capturing the underlying physics~\cite{Buoninfante:2024oxl}. Specifically, the teleological nature of event horizons makes them inherently undetectable in any finite-time experiment~\cite{Visser:2014zqa}. Similarly, spacetime singularities are generally thought to reflect our incomplete understanding of the behavior of spacetime and matter under extreme conditions~\cite{Crowther:2021qij}. Fundamentally, both concepts hinge on the notion of infinity: event horizons involve infinite time, while spacetime singularities correspond to infinite curvature, or equivalently, infinite energy and matter density. It is generally anticipated that these idealizations will be replaced by more appropriate concepts as the infinities are resolved into finite quantities.

As a result, standard black holes in general relativity are widely regarded as mathematical idealizations. While this concept has proven highly effective and has played an essential role in both mathematical and observational physics, the search for a less idealized, non-singular framework for black hole physics is clearly well-motivated. Paraphrasing Stephen Hawking, this ``suggests that black holes should be redefined as metastable bound states of the gravitational field''~\cite{Hawking:2014tga}. While a complete non-singular paradigm that can compete with the standard one remains elusive, decades of research have led to important partial results and insights. 

\enlargethispage{10pt}
This paper provides a comprehensive overview of this area of research, emphasizing recent developments and future prospects. The content was shaped through a structured, week-long focus program held at the Institute for Fundamental Physics of the Universe (IFPU) in Trieste from November 11th to 15th, 2024.
The workshop had an innovative structure in which each day was devoted to the study of a specific topic, including talks and a dedicated discussion session structured along three key questions that were communicated to all participants in advance. 

The structure of the paper reflects this approach, with each section (excluding introduction, conclusions, and glossary) corresponding to a thematic focus of the program and each subsection addressing one of the predefined questions. The discussion assumes familiarity with technical spacetime concepts, and readers may thus benefit from consulting the \hyperref[sec:glossary]{glossary of terms} appended at the end.


\section{Regular black holes and black hole mimickers}
\subsection{What is an adequate definition of being “non-singular”?}

The concept of singularities in general relativity is inherently multifaceted, influencing any discussion about what constitutes a “non-singular” framework~\cite{Bojowald:2007ky}. Nevertheless, a universal expectation for any physical theory is its ability to describe the evolution of physical trajectories smoothly, without abrupt interruptions, and to ensure that physical observables measured along these trajectories remain finite. This requirement naturally leads to two key aspects:

\begin{enumerate}
\item Completeness of physical trajectories: a fundamental theory that is predictive should not contain abrupt endings for physical trajectories. Non-physical trajectories (\emph{e.g.}, spacelike trajectories) may be incomplete without prejudice to the regularity of the theory.

\item Finiteness of physical observables: a fundamental theory that is predictive should yield finite values for physical observables along physical trajectories. Non-observable quantities may be divergent without prejudice to the regularity of the theory.

\end{enumerate}

Although these definitions are conceptually clear, their practical realization poses significant challenges and may not always be possible. The primary difficulty lies in the inherent complexity of determining the complete set of physical trajectories and observables for a given theory,\footnote[2]{Note that in some quantum gravity frameworks explicitly finding physical observables can be a nontrivial task, and actually a fundamental and foundational problem~\cite{Rovelli:1990ph}.} as this depends on factors such as the number, nature, and gravitational coupling of additional degrees of freedom as well as the gravitational dynamics itself. A secondary difficulty lies in establishing whether or not one's theory is truly fundamental. For instance the shock waves of classical hydrodynamics are purely mathematical singularities that are physically regularized by taking into account the modifications introduced by molecular dynamics. Given these limitations, a pragmatic approach is to adopt heuristic criteria, treating the completeness of metric geodesics and the finiteness of scalar curvature invariants as necessary conditions.\footnote[3]{It may be possible that, in specific frameworks or representations, the geodesics and curvature invariants associated with a given metric are not included in the set of physical trajectories and observables. This could make situations such as \emph{e.g.}~integrable singularities~\cite{Bonanno:2016dyv,Bonanno:2017zen,Adeifeoba:2018ydh,Borissova:2022mgd} compatible with these conceptual requirements.} 

\begin{figure}[!h]
\label{fig:intro}
\begin{center}
\includegraphics[width=1.0\linewidth]{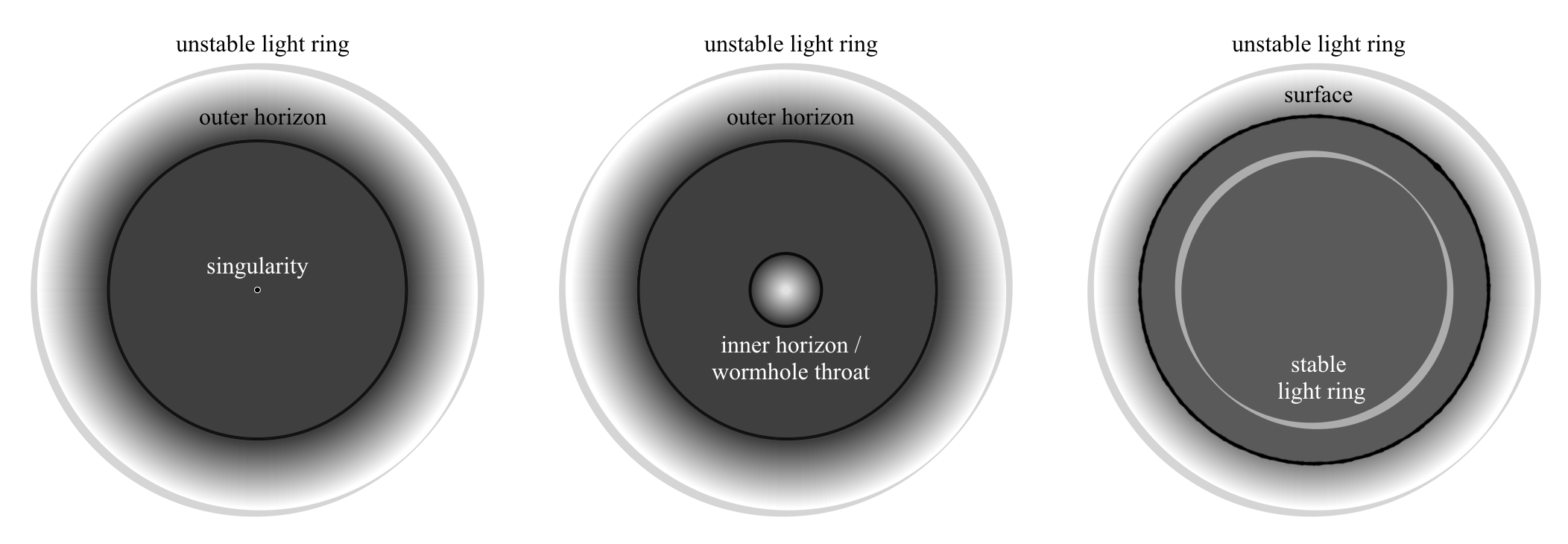}
\end{center}
\begin{center}
\caption{Pictorial representation of three
classes of spherically symmetric spacetimes, understood as constant-time snapshots, together with their distinctive features. \emph{Left:} a standard black hole has a trapped region (dark grey), with an outer boundary corresponding to an event horizon and containing a spacetime singularity (central dot). Standard black holes generically have an outer light ring (calligraphic circle in light grey). \emph{Center}: a regular black hole has a trapped region (dark grey), the outer boundary of which is an outer horizon (which may or may not become an event horizon in infinite time), and the inner boundary is an inner horizon that  encloses a non-singular core (light grey) or, alternatively, a spacelike wormhole throat. Regular black holes generically have an outer light ring (calligraphic circle in light grey). \emph{Right:} a black hole mimicker does not have a trapped region, but can be compact enough so that its interior mimics the properties of a trapped region. Black hole mimickers generically have outer and inner light rings (calligraphic circles in light grey).
}
\end{center}
\end{figure}

To illustrate the usefulness of these criteria, we can define classes of spacetimes satisfying them generically. For concreteness, we define two such classes below, to be contrasted with standard black holes in general relativity (see Fig.~\ref{fig:intro}):
\begin{itemize}
\item \textit{Standard black hole:} 
A spacetime that is an exact solution of the Einstein vacuum or electro-vacuum equations (or suitable generalizations). Standard black holes have a trapped region containing spacetime singularities and are bounded by an event horizon.
\item \textit{Regular black hole:} 
A spacetime that can be understood as a geometric deformation of some standard black hole. Regular black holes have trapped regions and outer horizons without spacetime singularities, which are replaced by regular cores with inner horizons or by spacelike wormhole throats. Trapped regions can last for an infinite or a finite time, and regular black holes have event horizons and Cauchy horizons in the former case.
\item \textit{Black hole mimicker:} 
A horizonless spacetime that can be understood as a geometric deformation of some standard black hole. Horizonless objects do not have a trapped region and can be ultracompact stars or traversable wormholes. The finite-redshift boundary of these objects will be generically referred to as the ``surface'' in the following. (Note that this ``surface'' does not have to be a ``hard'' surface, and that for objects with ``soft'' surfaces this notion might not be unequivocally defined).
\end{itemize}
The classes above are not intended to be complete. A more exhaustive geometrical classification of possible local regularizations of singularities can be found in references \cite{Carballo-Rubio:2019nel,Carballo-Rubio:2019fnb}. Furthermore, there are additional geometric possibilities, such as integrable singularities, that we will not further consider explicitly.

In general relativity, event horizons are intimately related to singularities, as encapsulated in the \emph{weak cosmic censorship hypothesis}~\cite{Penrose:1969pc,Wald:1992jst}. In the absence of spacetime singularities, event horizons are no longer required for the predictability of the theory, and thus are not necessary ingredients for the construction of regular black holes and black hole mimickers: the trapped region in regular black holes can disappear in finite time~\cite{Roman:1983zza,Hayward:2005gi}, while black hole mimickers have no trapped regions.
 
\subsection{Are there dynamical processes to go from regular black holes to black hole mimickers and the other way around?}

Yes, there are \emph{plausible} dynamical processes to go from regular black holes to black hole mimickers and the other way around:

\begin{enumerate}
    \item First, it is conceptually plausible for a black hole mimicker to eventually collapse into a black hole. Indeed, it has been argued that this is the generic outcome under the accretion of matter and energy if the mimicker's surface cannot sustain a spacelike evolution~\cite{Carballo-Rubio:2018vin} (see also~\cite{Carballo-Rubio:2018jzw,Chen:2019hfg}). Similar reasoning is frequently applied in thought experiments involving the merger of black hole mimickers of comparable size, often drawing on the ``hoop conjecture''~\cite{Cardoso:2019rvt}.

    \item Second, it is conceptually possible for a regular black hole to evolve dynamically into a black hole mimicker. The trapped region of a regular black hole could vanish in finite time due to various processes. Let us stress that pushing a classical black hole beyond extremality --- by overcharging or overspinning it --- would produce a naked singularity~\cite{Wald:1974hkz}, thereby violating the weak cosmic censorship conjecture~\cite{penrose1968battelle} (and indeed the feasibility of such a transition has been widely debated, see \emph{e.g.}~\cite{dafermos_bhio} for a recent critical review). In contrast, taking a macroscopic regular black hole beyond extremality, corresponds (at least at a purely geometrical level) to the formation of a black hole mimicker~\cite{Carballo-Rubio:2022nuj,Eichhorn:2022bbn,Eichhorn:2022oma}, a regular ultracompact object not implying any breakdown of the cosmic censorship conjecture, which does not need to be enforced in the regular case.\footnote[4]{The macroscopic black hole mimickers referred to in this section are not the Planckian remnants discussed in the context of the information loss problem~\cite{Susskind:1995da,Hossenfelder:2009xq,Ong:2024dnr}, but either extremal or super-extremal configurations with larger characteristic scales. Missing a complete dynamical framework for regular black holes, we are unable at the moment to determine if some of the arguments against the formation of extremal or super-extremal configurations existing in the literature~\cite{dafermos_bhio,Crisford:2017gsb,Aalsma:2020duv} would still apply.}

\end{enumerate}

   The physical mechanisms underlying such transitions are only partially understood, especially in the second case. A deeper understanding is essential to attribute physical meaning to such a mathematical transition and calculate the associated timescales. Proposed mechanisms for the transition of regular black holes to black hole mimickers include a dynamic growth of the core's size over time~\cite{Barcelo:2014cla,Carballo-Rubio:2022nuj}, which may be connected to inner horizon instabilities such as mass inflation and semiclassical backreaction~\cite{Barcelo:2020mjw,Barcelo:2022gii}, or the overspinning of the regular black hole~\cite{Eichhorn:2022bbn,Eichhorn:2022fcl}.

\subsection{Can a geometrodynamic theory generically cure singularities?
}

In the absence of evidence to the contrary, it is reasonable to assume that the tools of differential geometry and field theory are sufficient to describe non-singular scenarios. However, from a conceptual perspective, curing singularities need not be a universal feature, as there are at least two potential limitations to achieving this within such a framework:
\begin{enumerate}
    \item 
    A given theory might produce non-singular spacetimes from “reasonable” initial conditions while failing to do so in other cases, a behavior proposed to be essential for the self-consistency of gravitational theories~\cite{Horowitz:1995ta}.
    \item 
    Certain physically relevant situations may necessitate alternative approaches, such as discrete geometry, to adequately describe phenomena like the interior of black holes~\cite{Ashtekar:2005qt}.
\end{enumerate}
So far, there are specific proposals for geometrodynamic theories based on action principles that bypass both caveats~\cite{Ziprick:2010vb,Bonanno:2023rzk,Barenboim:2024dko,Frolov:2024hhe}. While these formalisms are still under development, there is a clear value in understanding the non-singular evolution of spacetimes that describe realistic processes, such as the gravitational collapse or merger of massive stars to form a regular black hole or black hole mimicker.

\section{Formation}
\subsection{What are the known physical mechanisms that could lead to the formation of regular black holes and black hole mimickers?}

Proposals for the formation of non-singular black holes typically require incorporating effects beyond general relativity, often inspired by semiclassical or quantum gravity considerations. These effects can be represented through an action that includes effective matter content violating energy conditions~\cite{Barcelo:2002bv} and/or higher-order curvature terms.

Several authors propose that new physics kicks in at the onset of horizon formation. For instance, in the gravastar proposal~\cite{Mazur:2001fv,Mottola:2023jxl,Visser:2003ge,Chirenti:2007mk} (see also~\cite{Jampolski:2023xwh}), classical black holes are replaced by black hole mimickers due to the occurrence of a phase transition that causes the breakdown of general relativity in the vicinity of any horizon~\cite{Chapline:2000en}, while a similar process takes place in the fuzzball proposal~\cite{Kraus:2015zda}, which is now associated with tunneling to black-hole microstates that prevents the formation of horizons~\cite{Mathur:2008kg,Bena:2013dka}. The main criticism of such proposals is that spacetime curvatures at the onset of horizon formation can be made arbitrarily low for large enough masses, and therefore that the existence of a universal mechanism to avoid horizon formation is unlikely~\cite{Barcelo:2009zz,Visser:2008rtf} (note, however, that whether curvature is always a good indicator of general relativity being in its strong-field regime is debatable). Existing studies of gravitational collapse in semiclassical gravity support this perspective~\cite{Barcelo:2007yk,Barcelo:2009tpa}, although extremal configurations seem to act as amplifiers of quantum effects~\cite{Horowitz:2023xyl}.

The other regime in which new physics has been proposed to appear is when Planckian densities are reached in the collapsing matter. This idea traces back to the works of Sakharov and Gliner, who proposed that the de Sitter geometry naturally arises in the description of matter at such extreme densities~\cite{Gliner:1966ebg,Sakharov:1966ads}. Independently, Bardeen introduced a spacetime model featuring a trapped region while remaining geodesically complete~\cite{Bardeen:1968qtr}. Subsequent contributions by Dymnikova~\cite{Dymnikova:1992ux}, Hayward~\cite{Hayward:2005gi}, and others established the concept of regular black holes, where a repulsive force stabilizes the core. This repulsive force can stem from various origins, and numerous frameworks with different physical motivations have yielded regular black hole geometries in distinct regimes. Examples include frameworks based on the weakening of gravitational interaction~\cite{Bonanno:2000ep}, spacetime discreteness~\cite{Modesto:2005zm}, bounded curvature invariants~\cite{Frolov:2022fsl}, semiclassical effects~\cite{Abedi:2015yga} or higher-curvature corrections~\cite{Frolov:1981mz,Holdom:2002xy,Koshelev:2024wfk}. In dimensions strictly greater than 4, it has been explicitly shown that regular black holes arise as vacuum solutions when considering infinite towers of higher-curvature corrections~\cite{Bueno:2024dgm}. Recent variations even explore core regions that are asymptotically empty --- asymptotically Minkowski cores~\cite{Simpson:2019mud,Berry:2020ntz,Simpson:2021dyo,Simpson:2021zfl}. Hence, there seems to be no shortage of physical effects that can lead to a regular black hole. However, most of these arguments have been focused on idealized situations such as static or spherically symmetric spacetimes, thus leaving unaddressed important dynamical questions, some of them explored below.

\subsection{What are the most promising formalisms for the study of dynamical aspects?}

The literature studying the dynamical formation of regular black holes and black hole mimickers is relatively limited, especially beyond spherical symmetry.
Several models leading to the formation of spherically symmetric regular black holes and inspired by quantum gravity approaches have been explored in the literature~\cite{Berry:2021hos}. In loop quantum gravity, the existence of a minimum area leads to regular solutions~\cite{Modesto:2005zm,Gao:2017ihf,Ashtekar:2018lag,Alesci:2019pbs,Kelly:2020lec,Husain:2021ojz,Husain:2022gwp,Ashtekar:2023cod}. In asymptotic safety, phenomenological
models have been built that include the weakening of the gravitational
interaction and take the form of regular black holes~\cite{Bonanno:2000ep,Torres:2014gta,Torres:2015aga,Bosma:2019aiu,Platania:2019kyx,Bonanno:2019ilz,Eichhorn:2022bgu,Platania:2023srt}, though in some cases (weaker) singularities do survive~\cite{Casadio:2010fw,Bonanno:2016dyv,Bonanno:2017zen,Adeifeoba:2018ydh,Borissova:2022mgd}. More recently, there has been a proposal for equations of motion encapsulating an energy dependence of the gravitational constant inspired by the asymptotic safety programme, which can describe the formation of regular black holes (and black hole mimickers in some regimes)~\cite{Bonanno:2023rzk}.

Numerical simulations of 2-dimensional regular black holes, based on a generic dilaton gravity action, have been reported in references~\cite{Ziprick:2010vb,Taves:2014laa}. These 2-dimensional simulations can be embedded as the spherically-symmetric sector of 4-dimensional spacetimes to provide hints about these more realistic situations. Some explicit examples of such an embedding have been studied for gravitational theories defined in dimensions strictly greater than 4~\cite{Colleaux:2019ckh,Bueno:2024eig,Bueno:2024zsx}. Later simulations~\cite{Barenboim:2024dko} have uncovered the formation of a sequence of trapped and anti-trapped regions, which was anticipated in a previous proposal for non-singular gravitational collapse~\cite{Barcelo:2014cla,Barcelo:2015noa}.

Existing effective actions can describe the dynamics of low-compactness black hole mimickers, such as boson stars~\cite{Guzman:2009zz,Siemonsen:2024snb} and, more recently, black shells~\cite{Pretorius:1997wr,Giri:2024cks}. However, more compact objects ---  which would serve as better black hole mimickers --- have eluded such treatments. Arguments for their potential formation remain partial and model-dependent, drawing from semiclassical physics~\cite{Barcelo:2007yk,Barcelo:2009tpa}, as well as proposals like gravastars~\cite{Mazur:2001fv,Mottola:2023jxl,Visser:2003ge,Chirenti:2007mk} and fuzzballs~\cite{Mathur:2008kg,Bena:2013dka}.

The Lorentzian path integral formalism can also provide valuable insights into the solution space of theories extending beyond general relativity. Finite-action configurations that are not suppressed in the Lorentzian path integral possess some non-zero  probability of formation, although this approach does not provide much information  about the detailed mechanisms of their formation. Research leveraging on the Lorentzian path integral suggests that higher-order curvature terms are crucial for resolving singularities~\cite{Borissova:2020knn}, possibly requiring an infinite series of such terms~\cite{Borissova:2023kzq}. Furthermore, these studies have placed constraints on the types of regular black holes that can emerge in the absence of matter~\cite{Knorr:2022kqp}.

\subsection{What are the open issues?}

The existence of action principles leading to the formation of regular black holes, in particular the phenomenological proposal in~\cite{Ziprick:2010vb} and the asymptotic-safety-inspired proposal in~\cite{Bonanno:2023rzk}, is an important step towards understanding the dynamical aspects of these objects. Numerical simulations of the formation of regular black holes are now available~\cite{Barenboim:2024dko}, which makes it reasonable to expect that the coming years will see further developments and refinements. The gap that needs to be filled to match the machinery and numerical insights gained for standard black holes is large, but these are encouraging signs that it may soon shrink.

Black hole mimickers are lagging even further behind, especially in the high-compactness limit in which these objects can better resemble standard black holes. Even when specific mechanisms plausibly leading to their formation have been identified, these mechanisms are most often not embedded in a coherent framework that allows for the exploration of their dynamics in situations that are not highly idealized (such as static configurations). An open issue is the possible formation of black hole mimickers as a result of the internal evolution of regular black holes~\cite{Barcelo:2020mjw,Barcelo:2022gii,Carballo-Rubio:2022nuj}, which goes hand in hand with the study of the dynamics of the latter. 

\section{Instabilities}
\subsection{What are the known instabilities associated with different spacetime features?}

There are several potential instabilities associated with different features of non-singular spacetimes:
\begin{itemize}
    \item \textbf{Event horizons and outer horizons:} Event horizons are unstable due to the presence of Hawking radiation~\cite{Hawking:1974rv,Hawking:1975vcx}, which also applies to slowly-evolving outer horizons~\cite{Barcelo:2010pj}. In view of the extremely long timescale of evaporation for macroscopic Schwarzschild black holes, $T\propto T_\mathrm{Planck} (M/M_\mathrm{Planck})^3$, some authors prefer to rephrase the discussion in terms of Hawking meta-stability~\cite{Barcelo:2010pj}. The standard temperature law for Hawking radiation is modified for regular black holes, with the Hawking temperature increasing until it reaches a maximum, and decreasing afterwards until approaching zero for the configuration in which the inner and outer horizons merge together. Within the purely radiative model, the final remnant state is reached only asymptotically in an infinite amount of time~\cite{Carballo-Rubio:2018pmi}.
    
    \item \textbf{Cauchy horizons and inner horizons:} Cauchy horizons are unstable due to the presence of the mass inflation instability~\cite{Poisson:1989zz,Poisson:1990eh,Bonanno:1994ma,Bonanno:1994qh,Brown:2011tv,Frolov:2017rjz,Carballo-Rubio:2018pmi,Mcmaken:2021isj,Visser:2024zkx}, which also applies to slowly-evolving inner horizons~\cite{Carballo-Rubio:2024dca}, both against classical and semiclassical perturbations~\cite{Barcelo:2020mjw,Barcelo:2022gii,McMaken:2024fvq}. An additional semiclassical instability at the level of the renormalized stress-energy tensor has been reported to appear for Cauchy horizons whenever the (squared) surface gravities of event and Cauchy horizons do not match, $\kappa^2_\mathrm{in} \neq \kappa^2_\mathrm{out}$~\cite{McMaken:2023uue}.
    While only discussed in specific models, objects with inner horizons might be affected by  kink instabilities~\cite{Maeda:2005yd} and (angular) Laplacian instabilities~\cite{DeFelice:2024seu}.

    \item \textbf{Extremal horizons:} Configurations with extremal horizons arise as intermediate possibilities between regular black holes and black hole mimickers~\cite{Carballo-Rubio:2022nuj} and have been proposed as viable configurations that do not suffer from the instabilities typical of these two classes of objects~\cite{DiFilippo:2024spj}.   However, extremal horizons have their own instabilities. For instance the Aretakis instability is a purely classical effect operating at the level of the wave equation~\cite{Aretakis:2010gd,Aretakis:2011ha,Aretakis:2011hc,Aretakis:2013dpa, Aretakis:2012ei,Angelopoulos:2018uwb,Aretakis:2023ast}. There are also semiclassical arguments for the instability of extremal horizons~\cite{Anderson:1995fw,Anderson:2000pg,Marolf:2010nd}, though perturbative analyses of semiclassical effects may not {necessarily} yield large corrections~\cite{Lowe:2000ka,Anderson:2000qu}.

    \item \textbf{Surfaces:} Physical surfaces (or any other structure replacing outer horizons, such as wormhole throats) of black hole mimickers must evolve either in a non-causal or non-local way to avoid the formation of outer horizons in the presence of accretion~\cite{Carballo-Rubio:2018vin} (see also~\cite{Chen:2019hfg,Nielsen:2005af}). This can therefore be understood as an instability of these configurations against accretion.

    \clearpage
    \item \textbf{Horizonless ergoregions:} Black hole mimickers that are rotating sufficiently fast are potentially unstable, as the absence of horizons makes ergoregions unstable against linear perturbations~\cite{Cardoso:2007az,Chirenti:2008pf,Maggio:2018ivz,Franzin:2022iai}. This linear instability might be quenched by absorption~\cite{Maggio:2017ivp}, or by a very long internal light-crossing timescale~\cite{Brustein:2024gia}, which effectively mimics absorption. 

    \item \textbf{Stable light rings:} Linear perturbations around stable light rings decay extremely slowly~\cite{Keir:2014oka}, which has been conjectured to result in a nonlinear instability~\cite{Cardoso:2014sna,Cunha:2017qtt} (see also ~\cite{Cunha:2022gde,Franzin:2023slm,Guo:2024cts}). The presence of stable light rings is generic for compact enough objects~\cite{Cunha:2017qtt,DiFilippo:2024ddg}. However, whether this results in a nonlinear instability, and the associated timescale, is still unclear (addressing this point may require a better understanding of self-gravity effects in light-rings; see \cite{DiFilippo:2024poc} for recent developments for unstable light rings).  A recent study points out that the leading effect is the development of turbulent behavior~\cite{Benomio:2024lev}. Recent numerical simulations of the merger of boson stars into a boson star with a stable light ring have observed no trace of this instability for the timescales explored in these simulations~\cite{Siemonsen:2024snb}.
    
    \item \textbf{Wormhole throats:} 
    The stability of wormholes~\cite{Morris:1988cz,Morris:1988tu,Visser:1989kh,Visser:1989kg,Cramer:1994qj,Boonserm:2018orb} has been studied less extensively compared to other classes of spacetimes. Notably, when a (spacelike) wormhole throat is located within an outer horizon~\cite{Simpson:2018tsi,Lobo:2020ffi,Franzin:2021vnj} (a geometry sometimes called a ``black bounce''), no instabilities have been identified, possibly setting this scenario favorably apart from other proposals. Nonetheless, wormhole throats stand out among the aforementioned alternative structures due to their reliance on a topology change for formation. This requirement introduces profound conceptual challenges and may even preclude the formation of wormholes entirely within the framework of purely geometrodynamic theories, \emph{i.e.}~not resorting to summing over topologies or quantum-gravitational transitions.
\end{itemize}

Finally, while it is remarkable that we can deduce the presence of instabilities without knowing the full dynamical field equations, it is clear that we can only have a partial understanding without the knowledge of the full theory. The study of the mechanisms described in this section must be complemented with fully nonlinear computations in specific models.

\subsection{Are instabilities shortcomings of the models, 
or are they informative regarding dynamics?}

While instabilities often signal potential issues with a model, they can also serve as valuable guiding principles for exploring the landscape of non-singular geometries. In particular, unstable configurations can be viewed as at most transient phases in a gravitational collapse, ultimately leading to the system settling into a stable configuration. Seen from this perspective, instabilities might be crucial in shaping the evolution of the collapsing geometry and in selecting viable non-singular long-living ground states for it.

\begin{figure}[!ht]
\label{fig:instabilitymap}
\begin{center}
\includegraphics[width=0.6\linewidth]{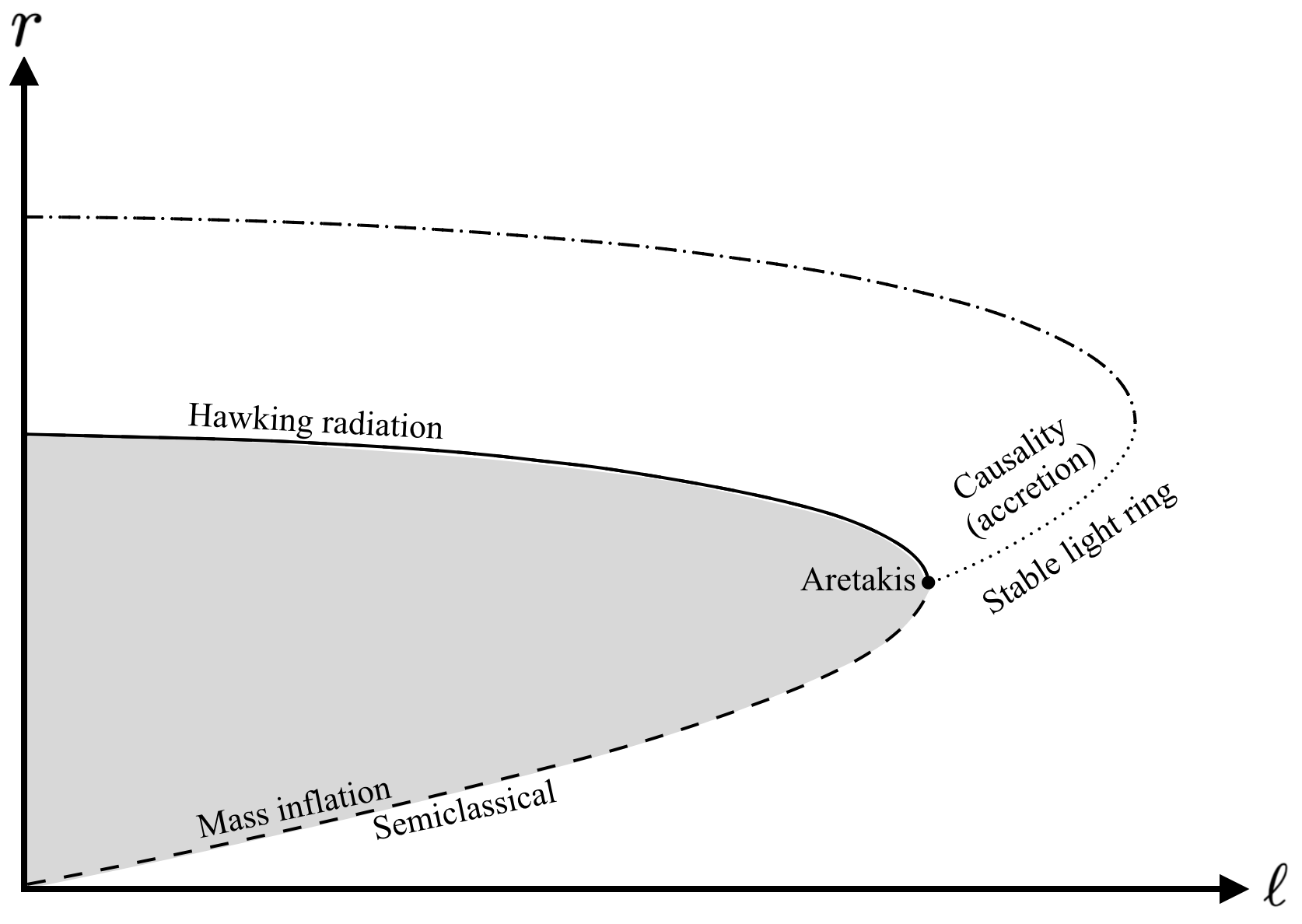}
\end{center}
\begin{center}
\caption{Schematic representation of the known instabilities for families of spherically symmetric spacetimes interpolating between regular black holes and black hole mimickers~\cite{Carballo-Rubio:2022nuj}. Each value of $\ell$ on the horizontal axis indicates a spacetime of the family, with the location of its characteristic features as a function of the radial coordinate $r$ being plotted on the vertical axis. Unstable light rings are indicated by the dashdotted line, stable light rings by the dotted line, outer horizons by the solid line, inner horizons by the dashed line and extremal horizons by the bullet point. Trapped regions are indicated by the shaded area.}
\end{center}
\end{figure}

Formal instabilities are always (at least partially) informative (see Fig.~\ref{fig:instabilitymap}). In some cases, they suggest that certain configurations are transient by nature. If an instability signals progression toward a satisfactory equilibrium, it is often seen as a feature rather than a flaw. However, this interpretation depends heavily on the specific instability under consideration.
In other cases, instabilities highlight the limitations of specific models. For instance, they might highlight the breakdown of the description of spacetime in terms of Lorentzian effective metrics, which may thus require to deal with averaged or coarse-grained behaviours, rather than rely on idealized continuous fields.

Ultimately the various known instabilities, be they rapid or slow, seem to strongly restrict the number of allowed equilibrium possibilities. 
Of course this also depends on the timescale of the instability. For instance, even if we have included Hawking radiation in the list of instabilities, as it implies that spacetimes with non-extremal event or outer horizons are at best meta-stable, and cannot describe the true ground state of gravitational collapse, the extremely long associated timescale implies that this effect is not observationally relevant for astrophysical black holes. 
For some instabilities  the relevant timescales are known, in other cases the relevant timescales are not particularly well understood.

\clearpage
Thus, it is clear that there are issues at play that require a much deeper understanding, in particular regarding the relevant timescales and the dynamical evolution triggered by these instabilities. These questions remain unanswered for most of the known instabilities. In fact, a quick look at Fig.~\ref{fig:instabilitymap} suggests that there would be no true ground state in a non-singular paradigm, which is compatible with the statement that, in such a paradigm, black holes would be defined as metastable bound states.

\subsection{What are the differences between the dynamics of 
classical versus semiclassical/quantum perturbations?}

There is a significant difference between detecting an instability (whether classical or semiclassical or quantum) at the linearized level, and acquiring a full understanding of the relevant dynamics ---  unfortunately it is the full dynamics that is needed to probe nonlinear stability, and that is exactly where the various competing timescales come into play. 

A particularly interesting feature of these instabilities is that some of them are purely classical, whereas others are intrinsically semiclassical. That some of these phenomena involve Planck's constant $\hbar$ while others do not is ultimately (simply by dimensional analysis) the reason why the relevant timescales can differ so much. 

Even for instabilities that manifest for both classical and semiclassical perturbations, such as mass inflation, the corresponding backreaction is generically different~\cite{Barcelo:2022gii} due to renormalized stress-energy tensors generically violating energy conditions~\cite{Barcelo:2002bv}. Semiclassical effects could therefore be anti-gravitational, offering the possibility of opening up trapped regions~\cite{Barcelo:2020mjw}. The importance of including rotation to properly understand semiclassical backreaction in regular black holes has been demonstrated recently~\cite{Klein:2024sdd,McMaken:2024fvq}.

Finally, although one can usually (at least conceptually) control the source of the classical instability and sometimes also ``turn it off”, the same is not true for semiclassical effects. These effects are usually sourced by quantum fluctuations, which cannot be controlled as easily as classical sources.

\section{Observational signatures}
\subsection{What are the different observational channels?}
New observational effects are associated with the modifications of the defining elements of standard black holes, namely event horizons and spacetime singularities. Replacing spacetime singularities with regular cores generally results into modifications of the spacetime curvature (equivalently, the gravitational field) of these objects that decay far away from the regular core. Replacing event horizons by outer horizons or physical surfaces can change the appearance of these objects as seen from outside, due to the interior region being disconnected causally only for a finite time or never disconnected. Both can lead to observable consequences in the gravitational and electromagnetic wave windows~\cite{Carballo-Rubio:2018jzw,Cardoso:2019rvt}. 

\begin{figure}[htbp]
    \centering
    \includegraphics[width=0.8\linewidth]{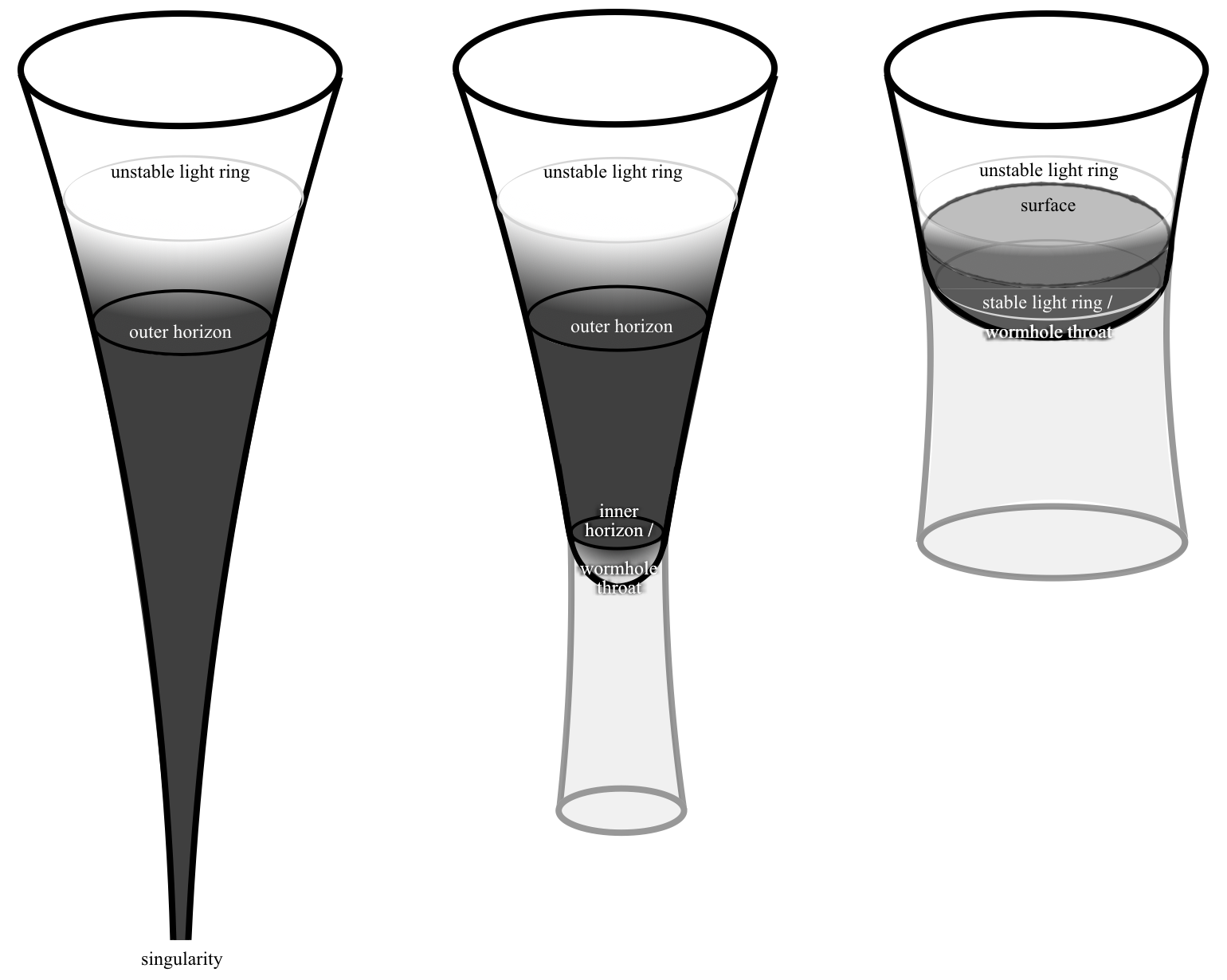}
    \caption{Standard black holes have an infinite gravitational well (diagram on the left), while both regular black holes and black hole mimickers have finite gravitational wells (middle and right figures, respectively), the difference between the two being the depth of the well. The shared feature of having deep gravitational wells (well below light rings) allows these objects to have a similar phenomenology. Trapped regions are marked in dark grey and light rings in calligraphic circles.}
    \label{fig:old-fig-1}
\end{figure}

Modifications of the spacetime curvature from that of the Kerr metric around either regular black holes or black hole mimickers can be detected in experiments that are able to constrain the gravitational field to exquisite precision. The most promising channels are observations of matter and electromagnetic or gravitational waves around supermassive black holes, including Sgr A$^*$ at the center of our galaxy. This includes observations of orbiting stars~\cite{Psaltis:2015uza,Anson:2021yli} and pulsars~\cite{Liu:2011ae,Pfahl:2003tf}, light originated in accretion disks and captured in black hole images~\cite{Psaltis:2018xkc,Younsi:2021dxe}, and gravitational waves produced in extreme mass-ratio inspirals (EMRIs)~\cite{Barausse:2020rsu}.

The lack of event horizons can have different consequences, depending on whether these are replaced by outer horizons or physical surfaces. Therefore, we discuss them separately. Spacetimes with long-lived outer horizons can have the same phenomenology as spacetimes with event horizons for arbitrarily long time intervals. Hence, the main observable consequences in this case would be associated with the trapped region closing up and releasing any radiation stored, which could result in bursts of gravitational radiation~\cite{Barcelo:2015noa} with a possible quasi-periodic nature~\cite{Barcelo:2016hgb,Barcelo:2017lnx}. It has been argued that quantum effects can be amplified in extremal configurations, which may lead to new observational signatures~\cite{Horowitz:2023xyl}.

On the other hand, black hole mimickers do not have trapped regions and therefore their interiors are not causally disconnected from the exterior. This opens the possibility of electromagnetic and gravitational waves propagating and interacting with these interiors, possibly leading to new effects. For electromagnetic waves, propagation effects can lead to new photon rings in black hole images~\cite{Mazur:2015kia,Eichhorn:2022bbn} while interaction effects can also lead to new photon rings~\cite{Carballo-Rubio:2022aed,EventHorizonTelescope:2022xqj} as well as thermal emission due to the surface and interior being heated up~\cite{EventHorizonTelescope:2022xqj,Carballo-Rubio:2023fjj}. For gravitational waves, propagation effects have been proposed to result in gravitational wave echoes~\cite{Cardoso:2016rao,Cardoso:2017cqb} as well as tidal heating~\cite{Maselli:2017cmm,Datta:2019epe,Cardoso:2022fbq}.

\subsection{What is the current theoretical/observational status of each channel?}

Our understanding of different observational channels is limited by the lack of knowledge of the dynamics of regular black holes and black hole mimickers. Hence, channels that can be analyzed by relying on linearized (test-field) approximations are better developed. The following is a list of specific channels with a brief summary of its current status:
\begin{itemize}

\item The inspiral part of gravitational waveforms can be reliably described in a linearized framework, also for extensions of general relativity~\cite{Will:1994fb}. Extreme mass-ratio inspirals (EMRIs) are particularly interesting events due to the hierarchy of masses between the two objects coalescing~\cite{Berry:2019wgg}, and offer clear opportunities to test general relavitity~\cite{Cardenas-Avendano:2024mqp}. EMRIs are expected to be frequent sources for the Laser Interferometer Space Antenna (LISA)~\cite{Naoz:2023hpz}.

\item The second phase in gravitational waveform is the merger phase, which is generally constructed by using a combination of numerical, analytical and phenomenological approaches~\cite{Buonanno:2000ef,Hannam:2013oca,Blackman:2015pia}. Due to the inherent complexity of the problem and the lack of numerical information for theories beyond general relativity, analyses of the merger phase are scarce, though this issue has been analyzed in specific theories such as Einstein-scalar-Gauss-Bonnet gravity~\cite{Julie:2024fwy}.

\item Black hole spectroscopy is a powerful technique to test the predictions of general relativity for the ringdown part of waveforms~\cite{Dreyer:2003bv,Berti:2005ys,Chirenti:2016hzd,Brito:2018rfr}. Its main benefits are that predictions can be extracted in a linearized approximation (however, it has been argued that nonlinearities are important~\cite{Cheung:2022rbm,Khera:2023oyf}), also for extensions of general relativity~\cite{Cano:2023jbk,Miguel:2023rzp,Chung:2024ira}, and are robust against environmental effects~\cite{Spieksma:2024voy}. Open challenges include, aside from the issues regarding nonlinearities mentioned above, determining which effects beyond general relativity can be reliably extracted from data gathered by next-generation interferometers~\cite{Franzin:2023slm,Maselli:2023khq,Kehagias:2024yyp}. 

\item Gravitational wave echoes are not part of waveforms in general relativity, as this phenomenon requires an interior that is not causally disconnected from the exterior, and would thus be a genuine sign of physics beyond general relativity~\cite{Cardoso:2016rao,Cardoso:2017cqb}. Searches for echoes have been implemented without success to date~\cite{Westerweck:2017hus,Lo:2018sep,Uchikata:2019frs}. There are still significant theoretical uncertainties about this phenomenon, ranging from the shape of individual echoes~\cite{Wang:2018gin}, their separation~\cite{Vellucci:2022hpl,Arrechea:2024nlp} and overall amplitude, which has been argued to be negligible due to backreaction effects (which can only be captured beyond the linearized approximation)~\cite{Carballo-Rubio:2018jzw,Chen:2019hfg,Guo:2022umn}.

\item Black hole mimickers can have thermal properties, emitting electromagnetic radiation as a consequence. Infrared observations of supermassive black holes, in particular Sgr A$^*$ and M87$^*$, are incompatible with these sources having physical surfaces in thermal equilibrium with their accreting environments~\cite{Broderick:2005xa,Narayan:2008bv,Broderick:2009ph,Broderick:2015tda} (see also the more recent discussions~\cite{EventHorizonTelescope:2022xqj,Carballo-Rubio:2023fjj}). Compact enough black hole mimickers cannot be ruled out due to the extreme gravitational lensing that prevents a sizable fraction of radiation from escaping~\cite{Lu:2017vdx,Carballo-Rubio:2018jzw,Cardoso:2019rvt}. Open issues include going beyond the equilibrium assumption to understand the thermalization timescale for specific black hole mimickers, which is in principle unbounded~\cite{Carballo-Rubio:2023fjj}.

\item The existence and properties of a central depression in black hole images, known as the black hole shadow~\cite{EventHorizonTelescope:2019dse,EventHorizonTelescope:2019pgp,EventHorizonTelescope:2019ggy,EventHorizonTelescope:2022wkp}, are important observables to constrain the spacetime geometry~\cite{EventHorizonTelescope:2021dqv,EventHorizonTelescope:2022xqj}. Future upgrades of the EHT, such as the proposed next generation Event Horizon Telescope (ngEHT), would lead to quantitative improvements in the measurement of shadows ~\cite{Johnson:2023ynn,Ayzenberg:2023hfw}.

\item Photon rings in black hole images are essential observables to disentangle astrophysics from spacetime physics~\cite{Johnson:2019ljv,Gralla:2019drh,Gralla:2020srx,Tiede:2022grp,Kocherlakota:2023qgo}. Simple models of regular black holes (with only a single new-physics scale, the size of the regular core) have a larger relative separation of photon rings~\cite{Guerrero:2021ues,Eichhorn:2022oma}, while black hole mimickers may display additional rings~\cite{Olmo:2021piq,Eichhorn:2022bbn,Eichhorn:2022fcl,Carballo-Rubio:2022aed}. Being able to extract information of the number of rings and their separation in black hole images is an open problem. Future upgrades of the EHT (either Earth-based, with high sensitivity and a second, higher frequency) or space-based (with high enough sensitivity) can come close to resolving the photon rings in the most optimistic cases~\cite{Tiede:2022grp,Carballo-Rubio:2022aed,Carballo-Rubio:2023ekp,Johnson:2024ttr,Lupsasca:2024xhq}.
\end{itemize}

In general, current observations can be equally described by a degeneracy of different theories or approximation schemes, including general relativity. This suggests that we should focus at least some effort on developing null tests for each model.

\subsection{Which are the most promising channels in terms of future experiments?}

Both electromagnetic and gravitational waves are highly promising for testing deviations from general relativity. It is important to note that these channels probe quite distinct regimes. Black hole images provide information about objects in their “stationary” (metastable)  configuration, and can primarily probe the kinematics of regular black holes and black hole mimickers. Gravitational waves, on the other hand, explore the dynamics of these objects, potentially offering more insights into new physics. However, they are also more susceptible to uncertainties arising from unmodeled deviations in their dynamics. Because the two observational channels are affected by different types of uncertainties, combining information from both could be crucial.

Further motivation to make use of all channels simultaneously is that different channels currently probe black holes in very different mass regimes. Black-hole uniqueness theorems~\cite{Robinson:1975bv,Robinson:2004zz,Mazur:2000pn} may not carry out to frameworks beyond general relativity, and there are indeed examples of theories that can easily accommodate large deviations from standard black holes in the supermassive range, while deviations are absent in the stellar-mass range~\cite{Eichhorn:2023iab}. Given the current theoretical uncertainty about the regions in the parameter space spanned by the mass and spin of black holes in which new physics is expected to arise, it is important not to limit unnecessarily the exploration of different channels.

A final point to stress is that current theoretical understanding is limited by a lack of studies on nonlinear dynamics beyond general relativity or for objects other than black holes, with a few exceptions that have been covered above. Developing nonlinear waveforms from the coalescence of regular objects --- though potentially very model-dependent --- is now essential.

In counterpoint, numerical relativity techniques have improved significantly during the last few decades, and so have the number and quality of the predictions. While probably we cannot get analytical results in some ranges for some phenomena, we feel that it is nevertheless interesting to try obtain analytic results where it can be done, since analytical results allow us to get a better understanding of what is happening.

\section{Realistic astrophysical modeling}
\subsection{What are the most relevant astrophysical uncertainties?}

For gravitational wave emission, the primary uncertainties stem from unmodeled (or only partially modeled) astrophysical
effects, such as orbital eccentricity~\cite{Divyajyoti:2023rht}, spin-induced precession~\cite{Miller:2023ncs}, tidal interactions~\cite{Huang:2020pba}, microlensing~\cite{Nakamura:1997sw}, and viscosity~\cite{Bishop:2022kzq}. These effects could be comparable to deviations from general relativity in the merger of regular black holes and black hole mimickers, and thus may play a role in detecting such deviations. On the other hand, environmental effects are negligible for most situations, although they are of importance for tests of deviations from general relativity~\cite{Barausse:2014tra}.

In contrast, for black hole images, significant astrophysical uncertainties arise due to the environment, particularly the physics of the accretion disk~\cite{Kocherlakota:2022jnz}. There are large uncertainties about the physics of accretion disks, including their magnetization state~\cite{Dhruv:2024igk} and the role of non-thermal emission~\cite{Zhang:2024ddt}. The lack of a more precise understanding of the emission profile of accretion disks limits the possibilities of extracting black hole images for regular black holes and black hole mimickers using available ray-tracing codes (see \emph{e.g.}~the resources cited in~\cite{Lupsasca:2024wkp}). A significant effort focussed on modeling these effects is therefore essential. 

\subsection{Which ingredients do we need to go from theoretical to realistic modeling?} 

For gravitational wave observations, the main tasks are the extraction of waveforms in theories beyond general relativity, and the comparison with existing astrophysical uncertainties in order to assess their detectability. This type of analysis has been performed, for instance, for black hole spectroscopy~\cite{Maselli:2023khq}. As astrophysical effects are expected to be source-dependent, analyzing populations of events should help improving the detectability of physics beyond general relativity.

For black hole images, it would be necessary to construct realistic images properly taking into account the physics of accretion disks~\cite{Moriyama:2023nqj}, and then 
simulate observations, which only sample an
image sparsely in the Fourier plane~\cite{EventHorizonTelescope:2019dse,EventHorizonTelescope:2022wkp} (see also~\cite{Patel:2022acr}), which has an important impact in the image reconstruction procedure~\cite{EventHorizonTelescope:2019uob,EventHorizonTelescope:2022wok}. The data from simulated observations can be then compared with Event Horizon Telescope (EHT) data in order to determine whether images of regular black holes or black hole mimickers may provide a better fit than those of standard black holes in some situations. Modeling the physics of accretion disks can be sidestepped in two ways: 
\begin{enumerate}
 \item   
Accessing the photon rings of high enough order (ideally $n\geq2$), as the location at which the
emission originated is in practice irrelevant for these image components~\cite{Johnson:2019ljv,Gralla:2019drh}. However, current telescope array configurations are not sufficient to resolve photon rings~\cite{Gralla:2020srx,Tiede:2022grp}.
\item 
Using the concept of lensing bands~\cite{Gralla:2020srx,Paugnat:2022qzy} to marginalize over astrophysical uncertainties~\cite{Cardenas-Avendano:2023obg}. For two spacetimes in which the lensing bands of order $n$ do not overlap, an observation that detects
the photon ring of order $n$ with sufficient precision can rule out one of these two spacetimes. This procedure can be used to cast conservative constraints that can be improved upon adequate modeling of astrophysical uncertainties.
\end{enumerate}

\subsection{Are potentially observable signatures hidden by astrophysical uncertainties?}

Without a better assessment of the features of observable signatures, it is not possible to answer this question conclusively. However, existing analyses show that small enough observable signatures can indeed be hidden by astrophysical uncertainties.

For instance, it has been pointed out that corrections to general relativity must be large enough not to be degenerate with environmental effects and thus be detectable using gravitational wave observations~\cite{Barausse:2014tra}. For black hole images, the concept of lensing bands~\cite{Gralla:2020srx,Paugnat:2022qzy} can be leveraged to illustrate that potentially observable signatures can indeed be hidden by astrophysical uncertainties, as lensing bands that overlap for two different spacetimes indicates the degeneracy of spacetime and matter effects for photon rings.

\section{Conclusions and outlook}

The theoretical motivation to construct a non-singular paradigm of black hole physics, in which black holes are not described as the singular end state of gravitational collapse, but rather as regular and metastable bound states of the gravitational field, is sound. Moreover, the required mathematical machinery to develop such a new paradigm is reaching a mature stage, with concepts such as outer horizons providing suitable replacements for event horizons, and the classes of spacetimes that could result from singularity regularization being determined. There is also a clear understanding of the possible phenomenological opportunities that could be opened. The main missing link between these elements is dynamics, although there have recently been promising advances in this direction as well. Having a solid grasp of the dynamics in such a new paradigm would be transformative for the field.

On the other hand, there is a steady stream of incoming observational data --- primarily gravitational wave data and black hole imaging, with some hope for a few bits of optical data as well. Current (LIGO/Virgo/KAGRA, the EHT) and planned (LISA, TianQin, Taiji, Einstein Telescope, Cosmic Explorer, Black Hole Explorer) observational instruments suggest another several decades of future incoming data --- which would be extremely useful in not letting the theorists get too far off track. While the interplay with observations is definitively positive for theoretical research, it is important to remark that the conceptual motivation for a new paradigm of black hole physics stands on its own. As in the case of general relativity itself, the motivation for constructing new physical theories can be purely conceptual, and observational breakthroughs can be the consequence of such a process instead of its starting point. Overall, the future of the field of regular black holes and black hole mimickers looks healthy, and the next years will probably see critical advances in the field.

\clearpage
\section*{Glossary of terms}
\addcontentsline{toc}{section}{Glossary of terms}
\label{sec:glossary}

For the sake of clarity, we collect here some brief definitions that we will be using frequently and that are standard in the literature.

\begin{itemize}
\item \textit{Event horizon:} 
Boundary of the region from which outgoing light rays cannot escape to spatial infinity. Teleological. 
\item \textit{Spacetime singularity:}
Naive definition --- locus of infinite curvature. Most common definition --- associated to incomplete geodesics. More subtle definition --- some observable quantity is ill-behaved.
\item\textit{Trapped surface:} A closed, spacelike, 2-surface such that both expansions of ingoing and outgoing null geodesics are negative.
\item \textit{Trapped region:} 
A trapped region on a 3-dimensional hypersurface is the set of all the points on the hypersurface through which a trapped surface passes.
\item\textit{Anti-trapped surface:} A closed, spacelike, 2-surface such that both expansions of ingoing and outgoing null geodesics are positive.
\item \textit{Anti-trapped region:} 
A trapped region on a 3-dimensional hypersurface is the set of all the points on the hypersurface through which an anti-trapped surface passes. Until recently not expected to be directly relevant for astrophysical observations, nowadays considered possibly relevant as transient phases of gravitational collapse. 
\item \textit{Outer horizon:} A 3-surface at which the outgoing null congruence expansion is zero, the ingoing one is negative and the Lie derivative of the former along the ingoing congruence is negative (also called Future Outer Trapping Horizon). Positive surface gravity. Exponential peeling of outgoing light rays. Emits Hawking radiation.
\item \textit{Inner horizon:} A 3-surface at which the outgoing null congruence expansion is zero, the ingoing one is negative and the Lie derivative of the former along the ingoing congruence is positive (also called Future Inner Trapping Horizon). Negative surface gravity. Exponential pile-up of ingoing light rays. Subject to mass inflation. 
\item \textit{Extremal horizon:} 
Zero surface gravity. Formed by fine-tuned merger of outer and inner horizons, or more generally associated to degenerate horizons.
\item \textit{Cauchy horizon:}
Boundary of the globally hyperbolic region; boundary of the region for which one can define a Cauchy surface. Often a static/stationary inner horizon.
\item \textit{Light ring (spacetime feature):}
Circular photon orbit. Typically equatorial.
\item \textit{Photon ring (black hole image):}
The visible/radio effect of light from the accretion disk, (or other background source), being focused by the gravitational well of the central object, resulting in an infinite sequence of ring-like features in the black hole image indexed by the number $n$ of photon orbits around the black hole. 
\item \textit{Lensing band:}
The photon ring of order $n$ in a black hole image must lie within a bounded region, known as the lensing band of order $n$. Uniquely determined by the spacetime geometry, become exponentially smaller with increasing $n$.
\end{itemize}
%


\section*{Acknowledgements}

The workshop organizers are thankful to IFPU for financial support of the focus week program ``Towards a Non-singular Paradigm of Black Hole Physics'' which inspired this publication. RCR acknowledges financial support provided by the Spanish Government through 
the Grant No.~PID2023-149018NB-C43 (funded~by MCIN/AEI/10.13039/501100011033), by the Junta de Andaluc\'{\i}a 
through the project FQM219 and from the Severo Ochoa grant 
CEX2021-001131-S funded by MCIN/AEI/ 10.13039/501100011033. FDF acknowledges financial support from the PRIMUS/23/SCI/005 and UNCE24/SCI/016 grants by Charles University, and the GAR 23-07457S grant from the Czech Science Foundation.
MV's travel and participation in the workshop was supported by a Victoria University of Wellington grant, by SISSA, and by INFN (Trieste).
\bigskip
\hrule
\bigskip
\noindent 
CB financial support was provided by the Spanish Government through 
the Grants No. PID2020-118159GB-C43 and PID2023-149018NB-C43 (funded 
by MCIN/AEI/10.13039/501100011033), by the Junta de Andaluc\'{\i}a 
through the project FQM219 and from the Severo Ochoa grant 
CEX2021-001131-S funded by MCIN/AEI/ 10.13039/501100011033. The Center of Gravity is a Center of Excellence funded by the Danish National Research Foundation under grant No. 184. PMM is supported by the project PID2022-138263NB-I00 funded by MICIN/AEI/10.13039/501100011033 and by ERDF/EU.
The research of FDP and AP is supported by a research grant (VIL60819) from VILLUM FONDEN.

\clearpage
\newcommand{\newblock}{}
\bibliographystyle{utphys}
\bibliography{references}

\addcontentsline{toc}{section}{References}

\newpage

\markboth{}{}

\section*{Affiliations:}

\addcontentsline{toc}{section}{Affiliations}

\address{$^{1}$ Instituto de Astrof\'isica de Andaluc\'ia (IAA-CSIC),
Glorieta de la Astronom\'ia, 18008 Granada, Spain}
\address{$^{2}$ Center of Gravity, Niels Bohr Institute, Blegdamsvej 17, 2100 Copenhagen, Denmark}
\address{${^3}$ Institut f\"ur Theoretische Physik, Max-von-Laue-Str. 1, 60438 Frankfurt, Germany}
\address{${^4}$ SISSA, Via Bonomea 265, 34136 Trieste, Italy and INFN Sezione di Trieste}
\address{$^{5}$ IFPU - Institute for Fundamental Physics of the Universe, Via Beirut 2, 34014 Trieste, Italy}
\address{$^{6}$ Victoria University of Wellington, Wellington 6140, New Zealand \bigskip\bigskip}
\address{$^{7}$ Osservatorio Astrofisico di Catania, Istituto Nazionale di Astrofisica (INAF),
Catania, Italy}
\address{$^{8}$ Istituto Nazionale di Fisica Nucleare (INFN), Sezione di Catania,
Catania, Italy}
\address{$^{9}$ Perimeter Institute, 31 Caroline Street North, Waterloo, ON, N2L 2Y5, Canada}
\address{$^{10}$ Department of Physics and Astronomy, University of Waterloo,
200 University Avenue West, Waterloo, ON, N2L 3G1, Canada}
\address{$^{11}$ CENTRA, Departamento de F\'isica, Instituto Superior T\'ecnico - IST, Universidade de Lisboa - UL, Avenida Rovisco Pais 1, 1049 Lisboa, Portugal}
\address{$^{12}$ Niels Bohr Institute, Blegdamsvej 17, 2100 Copenhagen, Denmark}
\address{$^{13}$ Institut f\"ur Theoretische Physik, Universit\"at Heidelberg, Philosophenweg 16, 69120 Heidelberg, Germany}
\address{$^{14}$ Departamento de F\'isica Te\'orica and IPARCOS, Universidad Complutense de Madrid,
E-28040 Madrid, Spain}
\address{$^{15}$ Universit\'e Paris-Saclay, CNRS/IN2P3, IJCLab, 91405 Orsay, France}
\address{$^{16}$ Department of Mathematics and Physics, University of Mary, Bismarck, North Dakota 58504, USA}

\address{$^{17}$ Dipartimento di Fisica e Astronomia ``Ettore Majorana'', Universit\'a di Catania, Via S.~Sofia 64, 95123, Catania, Italy}

\address{$^{18}$ Dipartimento di Fisica, ``Sapienza'' Universit\'a di Roma and Sezione INFN Roma 1, Piazzale Aldo Moro 5, 00185, Roma, Italy}

\address{$^{19}$ Frankfurt Institute for Advanced Studies, Ruth-Moufang-Str.~1, 60438 Frankfurt, Germany}

\address{$^{20}$ School of Mathematics, Trinity College, 17 Westland Row, Dublin 2, Ireland}

\address{$^{21}$ Quantum Theory Center ($\hbar$QTC) $\&$ D-IAS, Southern Denmark University, Campusvej 55, 5230 Odense M, Denmark}

\end{document}